\RequirePackage{amsthm}
\documentclass[sn-chicago, pdflatex]{sn-jnl}

\usepackage{graphicx}%
\usepackage{multirow}%
\usepackage[mathlines]{lineno}
\usepackage{amsmath,amssymb,amsfonts}%
\usepackage{amsthm}
\usepackage{mathrsfs}
\usepackage[title]{appendix}
\usepackage{xcolor}
\usepackage{textcomp}
\usepackage{manyfoot}
\usepackage{booktabs}
\usepackage{algorithm}
\usepackage{algorithmicx}
\usepackage{algpseudocode}
\usepackage{listings}
\usepackage{etoolbox}

\newcommand*\linenomathpatch[1]{%
  \cspreto{#1}{\linenomath}%
  \cspreto{#1*}{\linenomath}%
  \csappto{end#1}{\endlinenomath}%
  \csappto{end#1*}{\endlinenomath}%
}
\newcommand*\linenomathpatchAMS[1]{%
  \cspreto{#1}{\linenomathAMS}%
  \cspreto{#1*}{\linenomathAMS}%
  \csappto{end#1}{\endlinenomath}%
  \csappto{end#1*}{\endlinenomath}%
}

\expandafter\ifx\linenomath\linenomathWithnumbers
  \let\linenomathAMS\linenomathWithnumbers
  \patchcmd\linenomathAMS{\advance\postdisplaypenalty\linenopenalty}{}{}{}
\else
  \let\linenomathAMS\linenomathNonumbers
\fi

\linenomathpatch{equation}
\linenomathpatchAMS{gather}
\linenomathpatchAMS{multline}
\linenomathpatchAMS{align}
\linenomathpatchAMS{alignat}
\linenomathpatchAMS{flalign}

\makeatletter
\patchcmd{\mmeasure@}{\measuring@true}{
  \measuring@true
  \ifnum-\linenopenaltypar>\interdisplaylinepenalty
    \advance\interdisplaylinepenalty-\linenopenalty
  \fi
  }{}{}
\makeatother

\usepackage{hyperref}
\usepackage{natbib}
\usepackage{url}

\newtheorem{proposition}{Proposition}[section]
\newtheorem{definition}{Definition}[section]

\newtheorem{remark}{Remark}[section]

\usepackage[acronym]{glossaries}
\makeglossaries

\newacronym{egm}{EGM}{Endogenous Grid Method}
\newacronym{ez}{EZ}{Epstein-Zin}
\newacronym{ti}{TI}{Time Iteration}
\newacronym{vfi}{VFI}{Value Function Iteration}
\newacronym{eis}{EIS}{Elasticity of Intertemporal Substitution}
\newacronym{crra}{CRRA}{Constant Relative Risk Aversion}

\begin{document}

\title[]{The Endogenous Grid Method for Epstein-Zin Preferences}\author*[1,2]{\fnm{Alan}    \sur{Lujan}         
        \email{alujan@jhu.edu}
    } 
\affil[1]{
        \orgdiv{}
        \orgname{}
        \orgname{Krieger School of Arts and Sciences, Johns Hopkins University, Washington, DC, USA}
        \orgaddress{
            \street{}
            \city{}
            \postcode{}
            \state{}
            \country{}}}
\affil[2]{
        \orgdiv{}
        \orgname{}
        \orgname{Econ-ARK}
        \orgaddress{
            \street{}
            \city{}
            \postcode{}
            \state{}
            \country{}}}
\abstract{
The endogenous grid method (EGM) accelerates dynamic programming by inverting the Euler equation. This paper derives an EGM algorithm for consumption-savings problems with Epstein-Zin preferences. The transformation $W = V^{1 -\rho}$ recasts the Epstein-Zin certainty equivalent as a power mean whose inverse exists in closed form, and the EGM update applies that inverse at each grid point. Because the inversion is pointwise, the construction does not depend on the dimensionality of the state space; this paper presents the one-asset case, and the same inversion combines naturally with multi-asset and life-cycle settings. The resulting algorithm requires no root-finding, runs one to two orders of magnitude faster than value function iteration, and improves accuracy by more than one order of magnitude; at equal accuracy the speedup over value function iteration is two to three orders of magnitude. In a benchmark calibration with risk aversion 10 and elasticity of intertemporal substitution 1.5, the method solves the infinite-horizon problem on 100 grid points in 13 milliseconds with mean $\log_{10}$ Euler error of -4.8, and the welfare cost of the approximation is below 0.1\% of permanent consumption.
}
\maketitle


\section{Introduction}

Epstein-Zin preferences \citep{epstein1989substitution, weil1989equity} separate risk aversion from intertemporal substitution and have become standard in macro-finance \citep{weil1990nonexpected}, particularly in asset pricing \citep{bansalyaron2004}. Building on the temporal lottery framework of \citet{krepsporteus1978}, they are theoretically well grounded: \citet{duffieepstein1992} establish existence and uniqueness in continuous time, \citet{ren2020dynamic} provide modern discrete-time foundations, and \citet{epstein1991substitution} develop their empirical implementation. Solving the resulting consumption-savings problem quickly matters for calibration and for estimation by simulated method of moments.

The endogenous grid method \citep{carroll2006method} is the standard fast solver for time-separable consumption-savings problems, used widely in structural estimation of precautionary-saving and liquidity-constraint models in the tradition of \citet{deaton1991saving}. Extensions handle multiple controls \citep{baraborrillasfernandez2007}, non-smooth and non-concave problems \citep{fella2014generalization}, occasionally binding constraints \citep{hintermaierkoeniger2010endogenous}, discrete-continuous choice \citep{iskhakov2017endogenous}, and preferences whose marginal utility has no analytical inverse \citep{hallengreen2024endogenous}.

This paper derives an \acrshort{egm} algorithm for consumption-savings problems with Epstein-Zin preferences and identifies the structural reason it works. That reason, the quasi-arithmetic-mean (QAM) form of the certainty equivalent, does not depend on the dimensionality of the state space. The one-asset case is presented here for clarity; the same QAM inversion combines naturally with multi-asset and life-cycle settings.

Epstein-Zin first-order conditions are not new to the computational literature. \citet{kaplanviolante2014} derive them for a two-asset consumption-savings problem in their supplemental material and solve them by numerical root-finding, and \citet{druedahljorgensen2017} note that their generalized \acrshort{egm} framework accommodates Epstein-Zin preferences (their footnote 29). The most closely related prior work is \citet{coeurdacier2020financial}, who solve Epstein-Zin consumption-savings problems by time iteration: consumption, value, and expected value are treated as simultaneous controls, and the resulting Euler system requires numerical root-finding at every grid point in every iteration. What these papers leave open is the defining \acrshort{egm} step itself, the closed-form Euler inversion that eliminates root-finding, and the structural reason such an inversion exists; this paper supplies both.

The transformation $W = V^{1 -\rho}$ turns the Epstein-Zin certainty equivalent into a power mean. Within the broader class of quasi-arithmetic means $\varphi^{ -1}(\mathbb{E}[\varphi(X)])$ axiomatized by \citet{kolmogorov1930notion} and \citet{nagumo1930klasse}, power means are distinguished by the closed-form expression for $\varphi^{ -1}$. \acrshort{egm} applies that inverse at each grid point: an exogenous grid of end-of-period assets and a single application of $\varphi^{ -1}$ deliver consumption directly, without root-finding. The resulting algorithm tracks two functions instead of one (consumption and transformed value) but otherwise maintains the standard \acrshort{egm} structure. The same construction extends to other recursive preferences whose certainty equivalent reduces to a power-, log-, or exponential-mean, including the risk-sensitive preferences of \citet{hansensargent1995}.

On a standard \citet{bansalyaron2004} calibration with 100 grid points, \acrshort{ez}-\acrshort{egm} solves the problem in 13 ms with mean Euler error -4.8 in $\log_{10}$ units, against 219 ms and -3.6 for time iteration (\acrshort{ti}) and 758 ms and -3.3 for value function iteration (\acrshort{vfi}). \acrshort{egm} is therefore roughly 60 times faster than \acrshort{vfi}, 17 times faster than \acrshort{ti}, and more than an order of magnitude more accurate than either. In accurate mode \acrshort{ti} catches up on accuracy but runs 145 times slower than \acrshort{egm}; \acrshort{vfi} is slower still and never closes the accuracy gap. At equal accuracy, where each method uses the grid size needed to deliver comparable Euler errors, \acrshort{egm} is 166 to 744 times faster than \acrshort{vfi}. Welfare costs of the \acrshort{egm} approximation are below 0.1\% of permanent consumption.

The rest of the paper develops these claims in order: the~model~section derives the transformation and the inverted Euler equation, the~algorithm~section turns the inversion into \acrshort{ez}-\acrshort{egm}, and the~comparison~section sets up the value function iteration and time iteration benchmarks. The~benchmark~section reports speed, accuracy, the Howard acceleration profile, and robustness across the elasticity of intertemporal substitution (\acrshort{eis}), and the~conclusion discusses extensions to other recursive preferences and to multi-asset settings.

\section{Model}\label{sec-model}

\begin{definition}[Consumption-Savings Problem with Epstein-Zin Preferences]\label{def-model}An agent solves
\begin{equation}
V(m, z) = \max_c \left[ (1-\beta) c^{1-\rho} + \beta \left( \mathbb{E}_{z'|z} \left[ V(m', z')^{1-\gamma} \right] \right)^{\frac{1-\rho}{1-\gamma}} \right]^{\frac{1}{1-\rho}}
\end{equation}
subject to $a = m - c \geq 0$ and $m' = Ra + y(z')$, where $m$ is cash-on-hand, $c$ consumption, $a$ end-of-period assets, $\beta$ the discount factor, $R$ the gross interest rate, $z$ the current income state, and $y(z')$ income as a function of the realized state $z'$.

\end{definition}The state $z$ follows an AR(1) process with transition probabilities $\pi_{k\ell} = \Pr(z' = z_\ell \mid z = z_k)$, so the expectation $\mathbb{E}_{z'|z}$ is taken over next-period states conditional on the current state. The parameter $\rho$ governs the elasticity of intertemporal substitution ($\text{EIS} = 1/\rho$), while $\gamma$ governs risk aversion.\footnote{This paper follows the original \citet{epstein1989substitution} parameterization. An alternative convention, common in the asset pricing literature (e.g., \citet{bansalyaron2004}), defines the \acrshort{eis} directly as $\psi = 1/\rho$ and writes the recursion with $\psi$ in place of $1/\rho$. The auxiliary parameter is then $\theta = (1 -\gamma)/(1 -1/\psi)$, which equals $(1 -\gamma)/(1 -\rho)$ in the present notation.} When $\rho = \gamma$, preferences collapse to standard constant relative risk aversion (\acrshort{crra}) expected utility.

The derivation below assumes $\rho \neq 1$ and $\gamma \neq 1$; the limiting cases use logarithmic transformations but the method is otherwise analogous. For the infinite-horizon problem, the return impatience condition $\beta R^{1 -\rho} < 1$, equivalently $(\beta R)^{1/\rho} < R$, ensures the value function is well-defined; it holds at every calibration considered below ($\beta R = 0.979$). Risk aversion does not enter the bound: the certainty equivalent is a power mean, monotone and homogeneous of degree one, so $\gamma$ shapes the aggregation across states but not the growth of value along feasible paths.\footnote{The bound is elementary: income is bounded and resources grow at most at rate $R$, so the value function scales at most linearly in resources, $W = V^{1 -\rho}$ grows at most like $R^{(1 -\rho)t}$ along any feasible path, and the discounted recursion sums geometrically when $\beta R^{1 -\rho} < 1$. \citet{epstein1989substitution} establish existence of the recursive utility functional over consumption programs with bounded growth; for existence, uniqueness, and convergence of value and policy iteration in the savings problem itself see \citep{ren2020dynamic, jaskiewicz2024stochastic}, and for the exact (necessary and sufficient) spectral-radius condition for the utility recursion given a consumption process see \citet{borovickastachurski2020}. The recursion also requires $V > 0$; this follows by induction from the terminal condition whenever consumption remains strictly positive. When $\rho > 1$ the transformed value $W = V^{1 -\rho}$ diverges as $m \to 0$, but this reflects $V \to 0$ at the boundary rather than a failure of finiteness: next-period resources satisfy $m' = Ra + y(z') \geq \min_{\ell} y(z_\ell) > 0$, so $V$ is finite and strictly positive at every $m > 0$, and the algorithm stores $V$ rather than $W$ partly for this reason (Remark~\ref{rem-storing-v}).}

\subsection{Transformation}

\begin{definition}[Transformed Value Function]\label{def-transform}The \textit{transformed value function} is $W(m, z) \equiv V(m, z)^{1 -\rho}$.

\end{definition}This transformation simplifies the certainty equivalent to a power mean.\footnote{When $\rho > 1$, the transformation $W = V^{1 -\rho}$ is decreasing, so maximizing $V$ is equivalent to minimizing $W$. The Euler equation characterizes the optimum in either case, and \acrshort{egm} inverts it directly; the~robustness~analysis verifies this numerically across $\rho$.}\footnote{Among power transformations $V \mapsto V^a$ with $a \neq 0$, only $a = 1 -\rho$ converts the Bellman equation in Definition~\ref{def-model} into an additive form. Raising both sides of Definition~\ref{def-model} to power $a$ produces a CES aggregator with outer exponent $a/(1 -\rho)$; only $a = 1 -\rho$ eliminates it, yielding the additive form in (\ref{eq-bellman-w}). Other power transformations are valid changes of variable but leave the outer exponent in place, frustrating the closed-form Euler inversion that follows.} \citet{meyergohde2019entropy} use the same transformation to analyze model uncertainty with recursive preferences.

\begin{definition}[Certainty Equivalent]\label{def-mu}Let the \textit{auxiliary parameter} $\theta \equiv (1 -\gamma)/(1 -\rho)$. The \textit{certainty equivalent} of next-period transformed value is
\begin{equation}
\mu(a, z) \equiv \left( \mathbb{E}_{z'|z} \left[ W(m', z')^{\theta} \right] \right)^{1/\theta},
\end{equation}
where $m' = Ra + y(z')$.

\end{definition}The case $\theta = 1$ (equivalently $\gamma = \rho$) recovers time-separable \acrshort{crra} expected utility; when $\gamma > \rho$, the agent prefers early resolution of uncertainty; when $\gamma < \rho$, late resolution (see \citet{backus2008recursive} for a textbook treatment).

\begin{remark}[Quasi-arithmetic mean structure]\label{rem-qam}The certainty equivalent $\mu$ takes the form of a quasi-arithmetic mean of $W$ with generator $\varphi(w) = w^{\theta}$:
\begin{equation}
\mu(a, z) = \varphi^{-1}\!\left(\mathbb{E}_{z'|z}\left[\varphi\bigl(W(m', z')\bigr)\right]\right).
\end{equation}
\citet{kolmogorov1930notion} and \citet{nagumo1930klasse} axiomatize this class. Power generators $\varphi(w) = w^{\theta}$ are distinguished within it because the inverse $\varphi^{ -1}(y) = y^{1/\theta}$ has closed form; arbitrary quasi-arithmetic generators typically do not. The \acrshort{egm} step in Proposition~\ref{prop-euler} below exploits this closed-form inverse.

\end{remark}Raising the Bellman equation in Definition~\ref{def-model} to the power $1 -\rho$ yields the transformed Bellman equation:
\begin{equation}
\label{eq-bellman-w}
W(m, z) = \operatorname*{opt}_c \left[ (1-\beta) c^{1-\rho} + \beta \mu(m-c, z) \right],
\end{equation}
where $\operatorname{opt}$ denotes maximization when $\rho < 1$ and minimization when $\rho > 1$: the transformation $W = V^{1 -\rho}$ is increasing in the first case and decreasing in the second, so optimizing $V$ corresponds to the opposite extremum of $W$ when $\rho > 1$. The first-order condition, and hence the \acrshort{egm} step, is identical in both cases.

The transformation succeeds because it converts the CES aggregator in Definition~\ref{def-model} into an additive structure. The original Bellman equation involves $V$ raised to fractional powers both inside and outside the expectation; the transformed equation (\ref{eq-bellman-w}) is additive, with the nonlinearity isolated in $\mu$. This additive structure admits clean differentiation, yielding an Euler equation that depends on $W$ and $c$ but can be inverted for $c$ in closed form.

The first-order condition equates the marginal utility of consumption to the marginal value of savings:
\begin{equation}
\label{eq-foc}
(1-\beta)(1-\rho) c^{-\rho} = \beta \frac{\partial \mu(a, z)}{\partial a}.
\end{equation}
The envelope theorem, combined with the first-order condition, gives
\begin{equation}
\label{eq-envelope}
\frac{\partial W(m, z)}{\partial m} = (1-\beta)(1-\rho) c(m,z)^{-\rho}.
\end{equation}

Differentiating $\mu(a, z)$ with respect to $a$ yields:
\begin{equation}
\label{eq-mu-prime}
\frac{\partial \mu}{\partial a} = R \cdot \mu(a,z)^{1-\theta} \cdot \mathbb{E}_{z'|z}\left[ W(m', z')^{\theta-1} \cdot \frac{\partial W(m', z')}{\partial m'} \right].
\end{equation}
Substituting (\ref{eq-mu-prime}) into the FOC (\ref{eq-foc}) and applying the envelope condition (\ref{eq-envelope}) to the next-period marginal value, the Euler equation becomes
\begin{equation}
\label{eq-euler}
c^{-\rho} = \beta R \cdot \mu(a, z)^{1-\theta} \cdot \Xi(a, z),
\end{equation}
where the \textit{marginal expectation} $\Xi$ collects the next-period covariation between transformed value and marginal utility,
\begin{equation}
\Xi(a, z) \equiv \mathbb{E}_{z'|z} \left[ W(m', z')^{\theta-1} \cdot c(m', z')^{-\rho} \right].
\end{equation}

Given $W(\cdot, z')$ and $c(\cdot, z')$ for all $z'$, both $\mu(a,z)$ and $\Xi(a,z)$ depend only on end-of-period assets $a$ and the current state $z$. Raising (\ref{eq-euler}) to the power $-1/\rho$ and isolating consumption yields the central result:

\begin{proposition}[Inverted Euler Equation]\label{prop-euler}The Euler equation (\ref{eq-euler}) for Epstein-Zin preferences can be inverted to yield
\begin{equation}
c(a, z) = \left( \beta R \cdot \mu(a, z)^{1-\theta} \cdot \Xi(a, z) \right)^{-1/\rho}.
\end{equation}

\end{proposition}Because the right-hand side depends only on $(a, z)$ and the next-period functions $W(\cdot, z')$ and $c(\cdot, z')$, this closed form enables \acrshort{egm}: given an exogenous grid over $a$, the formula delivers $c(a, z)$ directly and the budget constraint recovers $m = c + a$.

\begin{remark}[Pointwise inversion and state-space dimension]\label{rem-pointwise}The inversion in Proposition~\ref{prop-euler} is pointwise: at each $(a, z)$, $\varphi^{ -1}$ acts on a scalar argument. The argument therefore does not depend on the dimensionality of the state space. When $a$ is a vector of asset stocks, the grid becomes multi-dimensional but the inversion at each grid point remains the same scalar operation, so the QAM inversion is compatible with standard multi-dimensional \acrshort{egm} machinery; the~conclusion discusses specific combinations.

\end{remark}\begin{remark}[Limiting Cases]\label{rem-limiting}When $\rho \to 1$, the transformation becomes $W = \log V$ and the certainty equivalent becomes $\mu(a,z) = \frac{1}{1 -\gamma}\log \mathbb{E}_{z'|z}[\exp((1 -\gamma)W(m',z'))]$, yielding the risk-sensitive preferences of \citet{hansensargent1995}; \citet{tallarini2000risk} demonstrates the equivalence in business cycle models. When $\gamma \to 1$, the certainty equivalent of $V'$ becomes the geometric mean $\exp(\mathbb{E}_{z'|z}[\log V(m',z')])$; \citet{giovanniniweil1989} show that this case yields myopic portfolio allocation.\footnote{The logarithmic limit requires renormalization: formally, $\lim_{\rho \to 1}(V^{1 -\rho} - 1)/(1 -\rho) = \log V$ by L'Hôpital's rule. The Bellman equation and certainty equivalent formulae stated here are the limiting forms after this renormalization.}

\end{remark}\begin{remark}[Storing V instead of W]\label{rem-storing-v}The derivation uses $W = V^{1 -\rho}$, but numerical implementation is better served by storing and interpolating $V$ directly and converting to $W$ only when needed. When $\rho > 1$, the transformation maps small values of $V$ to large values of $W$, so $W$ becomes less stable for interpolation near the borrowing constraint. When $\rho < 1$, both are well-behaved. In either case $V$ is the natural object to store. The algorithm below carries $(c, V)$, converts $V \to W$ for the Euler step, and converts back after the value update. The same code path works for $\rho < 1$ (\acrshort{eis} $> 1$) and $\rho > 1$ (\acrshort{eis} $< 1$); although $\theta$ changes sign at $\rho = 1$, the Euler equation and certainty equivalent formulas remain valid.\footnote{See the~robustness~analysis for numerical verification across $\rho$.}

\end{remark}\subsection{The direct approach}\label{sec-direct}

The transformation is not the only route to a closed-form inversion.\footnote{I thank Jeppe Druedahl for correspondence that prompted this subsection.} Differentiating the Bellman equation in Definition~\ref{def-model} directly requires only the $V$-space certainty equivalent $\nu(a, z) \equiv \left( \mathbb{E}_{z'|z} \left[ V(m', z')^{1 -\gamma} \right] \right)^{1/(1 -\gamma)}$. The first-order condition is $(1 -\beta)\, c^{ -\rho} = \beta\, \nu^{ -\rho}\, \partial \nu/\partial a$ and the envelope condition is $\partial V/\partial m = (1 -\beta)\, c^{ -\rho}\, V^{\rho}$; each step parallels the transformed derivation above, and the problem remains a maximization for every $\rho$, since no decreasing transformation is applied. Combining the two conditions yields the Euler equation
\begin{equation}
\label{eq-euler-direct}
c^{-\rho} = \beta R \cdot \mathbb{E}_{z'|z}\left[ c(m',z')^{-\rho} \cdot \left( \frac{\nu(a,z)}{V(m',z')} \right)^{\gamma - \rho} \right],
\end{equation}
which inverts for $c$ in closed form just as Proposition~\ref{prop-euler} does, and is in fact the same equation: $W = V^{1 -\rho}$ and $\theta(1 -\rho) = 1 -\gamma$ give $\mu = \nu^{1 -\rho}$ and $W'^{\theta -1} = V'^{\rho -\gamma}$, so $\mu^{1 -\theta}\, \Xi = \nu^{\gamma -\rho}\, \mathbb{E}_{z'|z}\bigl[ V(m',z')^{\rho -\gamma}\, c(m',z')^{ -\rho} \bigr]$, which is the expectation in (\ref{eq-euler-direct}).

The closed-form inversion therefore does not depend on the change of variables, and both routes track the same two functions at the same cost. What the transformation contributes is structure: it identifies the certainty equivalent as a quasi-arithmetic mean whose generator has a closed-form inverse (Remark~\ref{rem-qam}), it separates the right-hand side of the Euler equation into the two independently precomputable expectations $\mu$ and $\Xi$ where the direct route carries the single ratio $\nu/V'$, and it connects directly to the limiting cases in Remark~\ref{rem-limiting}.

\section{Algorithm}\label{sec-algorithm}

One distinction from standard \acrshort{egm} emerges from the structure of Proposition~\ref{prop-euler}: the Euler equation requires both the policy function $c(\cdot, z')$ and the value function $W(\cdot, z')$ evaluated at next-period states. In standard \acrshort{crra} problems the Euler equation depends only on $c(\cdot, z')$, so \acrshort{egm} can iterate on the policy function alone. Here both functions must be iterated until they jointly converge.

This makes the algorithm a hybrid of \acrshort{egm} and the Howard policy iteration technique introduced by \citet{howard1960dynamic}. \acrshort{egm} provides the root-finding-free policy update, while tracking $W(\cdot)$ alongside $c(\cdot)$ ensures the value function remains consistent with the policy. Each iteration updates both functions, and convergence requires both to stabilize.

The algorithm proceeds as follows. Fix a grid of end-of-period asset values $\{a_j\}_{j=1}^J$ with $a_1 = 0$ and an exogenous grid of cash-on-hand values $\{m_i\}_{i=1}^I$. Initialize with a policy such as $c^{(0)}(m, z) = m$ and a monotonically increasing, concave $V^{(0)}$ (e.g., $V^{(0)} = c^{(0)}$); in practice the iteration converges to a fixed point from such initializations. For finite-horizon problems, the terminal condition is $c(m, z) = m$; for infinite-horizon, iterate until convergence.

\begin{algorithm}
\caption{EGM for Epstein-Zin}\label{alg:egm-ez}
\begin{algorithmic}[1]
\Require Grid of end-of-period assets $\{a_j\}_{j=1}^J$, exogenous grid $\{m_i\}_{i=1}^I$, initial guess $(c^{(0)}, V^{(0)})$
\Ensure Converged policy $c(\cdot, \cdot)$ and value $V(\cdot, \cdot)$
\State Given $(c^{(n)}, V^{(n)})$ from iteration $n$, for each income state $z_k$:
\For{each $a_j$}
    \State Compute $m'_{j\ell} = R a_j + y(z'_\ell)$ for all $z'_\ell$
    \State Interpolate $c^{(n)}(m'_{j\ell}, z'_\ell)$ and $V^{(n)}(m'_{j\ell}, z'_\ell)$; compute $W^{(n)} = (V^{(n)})^{1-\rho}$
    \State Compute $\mu_j = \left( \sum_\ell \pi_{k\ell} W^{(n)}(m'_{j\ell}, z'_\ell)^{\theta} \right)^{1/\theta}$
    \State Compute $\Xi_j = \sum_\ell \pi_{k\ell} \left[ W^{(n)}(m'_{j\ell}, z'_\ell)^{\theta-1} \cdot c^{(n)}(m'_{j\ell}, z'_\ell)^{-\rho} \right]$
    \State Invert Euler: $c_j = \left( \beta R \cdot \mu_j^{1-\theta} \cdot \Xi_j \right)^{-1/\rho}$
    \State Recover grid: $m_j = c_j + a_j$
\EndFor
\State Append $(0, 0)$ at the constraint
\State Interpolate $c^{(n+1)}(\cdot, z_k)$ from $\{(m_j, c_j)\}$ to $\{m_i\}$
\For{each $m_i$}
    \State Compute $a_i = m_i - c^{(n+1)}(m_i, z_k)$
    \State Interpolate $\mu(a_i, z_k)$ from $\{\mu_j\}$
    \State Update: $W^{(n+1)}(m_i, z_k) = (1-\beta) c^{(n+1)}(m_i, z_k)^{1-\rho} + \beta \mu(a_i, z_k)$
    \State Convert: $V^{(n+1)}(m_i, z_k) = (W^{(n+1)}(m_i, z_k))^{1/(1-\rho)}$
\EndFor
\State \textbf{Iterate} until $\|c^{(n)} - c^{(n-1)}\| < \varepsilon$
\end{algorithmic}
\end{algorithm}
Each iteration involves only interpolation, expectation, and the closed-form inversion in Proposition~\ref{prop-euler}. The cost is comparable to standard \acrshort{egm}; the only addition is tracking the value function alongside consumption. Convergence is monitored via $\|c^{(n)} - c^{(n -1)}\|$, the standard \acrshort{egm} stopping rule. Monitoring consumption alone suffices: once the policy stops changing, the value update reduces to the utility recursion for the consumption process the policy induces, which converges under the finite-value condition of the~model~section \citep{borovickastachurski2020}, so $W$ stabilizes once $c$ has. The accuracy of the resulting policy and value functions is verified directly through the Euler-error diagnostics in the~benchmark~section.

The borrowing constraint is handled by augmenting the endogenous grid with the point $(m, c) = (0, 0)$ to anchor interpolation; because $a_1 = 0$, the first endogenous point satisfies $m_1 = c_1$, so linear interpolation between $(0, 0)$ and $(m_1, c_1)$ reproduces $c = m$ exactly below $m_1$, and $\mu(0, z)$ is a grid value rather than an extrapolation. For typical calibrations the resulting grid is monotonic and interpolation proceeds directly. If non-monotonicity occurs (i.e., $m_j > m_{j+1}$ for some $j$ despite $a_j < a_{j+1}$), construct the consumption function via the upper envelope following \citet{iskhakov2017endogenous}. In the constrained region where $a = 0$ binds, $c(m, z) = m$ and $W(m, z) = (1 -\beta) m^{1 -\rho} + \beta \mu(0, z)$.

\begin{figure}[!htbp]
\centering
\includegraphics[width=0.7\linewidth]{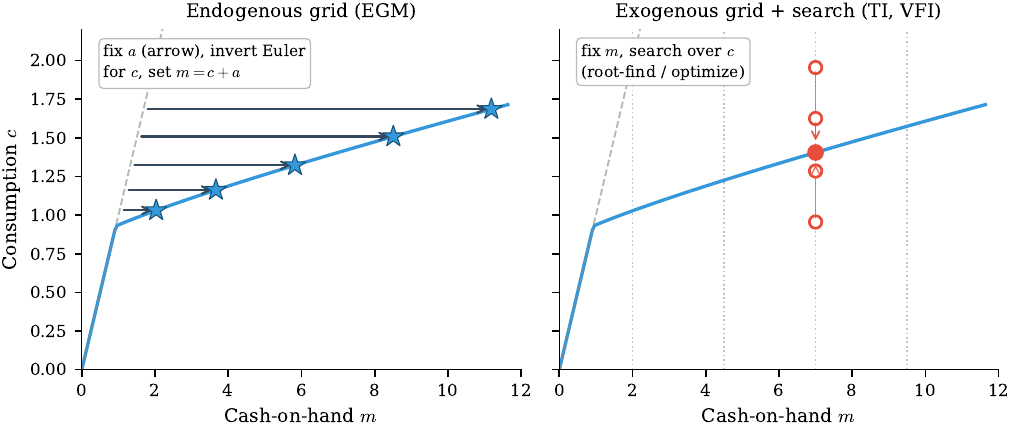}
\caption[]{Why \acrshort{ez}-\acrshort{egm} needs no root-finding. \textit{Left:} end-of-period assets $a = m - c$ are the horizontal gap between the policy and the $c = m$ line; fixing $a$ and inverting the Euler equation in Proposition~\ref{prop-euler} delivers consumption in closed form, so the endogenous grid point $m = c + a$ lands exactly on the policy. \textit{Right:} time iteration and value function iteration fix cash-on-hand $m$ and search over consumption at every grid point (bisection for \acrshort{ti}, golden-section for \acrshort{vfi}). The curve is the solved mean-income policy.}
\label{fig-mechanism}
\end{figure}

The contrast with search-based solvers is the source of the speed gains. Figure~\ref{fig-mechanism} places the two side by side: \acrshort{egm} fixes end-of-period assets and inverts the Euler equation in closed form, so every grid point lands on the policy without iteration, whereas time iteration and value function iteration fix cash-on-hand and search over consumption at each grid point.

\section{Comparison methods}\label{sec-comparison}

Value function iteration and time iteration, the standard dynamic programming approaches \citep{stokeylucasprescott1989recursive, judd1998numerical}, are the benchmarks. Both work with $V$ directly, so the relevant certainty equivalent is the $\nu$ of the~direct~approach, and both face an inherent tradeoff that \acrshort{egm} avoids: $\nu$ must be evaluated at candidate consumption values during search, either exactly at each candidate (\texttt{accurate} mode) or through interpolation of a precomputed asset-grid table (\texttt{fast} mode). Accurate mode preserves precision but incurs many $\nu$ evaluations per grid point; fast mode is cheaper but introduces approximation error. \acrshort{egm} has no such distinction because the endogenous grid places evaluation points exactly where the Euler equation is satisfied.

\subsection{Value function iteration}\label{sec-vfi}

\acrshort{vfi} finds optimal consumption by maximizing the Bellman equation directly,
\begin{equation}
V(m, z) = \max_c \left[ (1-\beta) c^{1-\rho} + \beta \nu(m-c, z)^{1-\rho} \right]^{1/(1-\rho)},
\end{equation}
where $\nu(a, z) = \left( \mathbb{E}_{z'|z}[V(Ra + y(z'), z')^{1 -\gamma}] \right)^{1/(1 -\gamma)}$ is the $V$-space certainty equivalent. Golden-section search evaluates $\nu$ at many candidate consumption values.

\begin{algorithm}
\caption{VFI for Epstein-Zin}\label{alg:vfi-ez}
\begin{algorithmic}[1]
\Require Exogenous grid $\{m_i\}_{i=1}^I$, asset grid $\{a_j\}_{j=1}^J$, initial guess $V^{(0)}$, mode $\in \{\texttt{accurate}, \texttt{fast}\}$
\Ensure Converged value $V(\cdot, \cdot)$ and policy $c(\cdot, \cdot)$
\State Given $V^{(n)}$ from iteration $n$
\If{mode = \texttt{fast}}
    \State Precompute $\nu_j = (\sum_\ell \pi_{k\ell} V^{(n)}(Ra_j + y(z'_\ell), z'_\ell)^{1-\gamma})^{1/(1-\gamma)}$ for all $a_j$
\EndIf
\For{each income state $z_k$ and each $m_i$}
    \State Define objective $f(c) = [(1-\beta) c^{1-\rho} + \beta \nu(m_i - c, z_k)^{1-\rho}]^{1/(1-\rho)}$
    \If{mode = \texttt{accurate}}
        \State Compute $\nu$ by interpolating $V^{(n)}$ at $m' = R(m_i - c) + y(z')$, then taking expectation
    \Else
        \State Interpolate $\nu$ from precomputed $\{\nu_j\}$ at $a = m_i - c$
    \EndIf
    \State Maximize $f(c)$ via golden-section search to get $c^{(n+1)}(m_i, z_k)$ and $V^{(n+1)}(m_i, z_k)$
\EndFor
\State \textbf{Iterate} until $\|V^{(n)} - V^{(n-1)}\| < \varepsilon$
\end{algorithmic}
\end{algorithm}
\subsection{Time iteration}\label{sec-ti}

Time iteration (\acrshort{ti}), introduced by \citet{coleman1991equilibrium}, works directly with the Euler equation rather than the Bellman equation. Like \acrshort{egm}, \acrshort{ti} exploits the first-order condition, but it fixes an exogenous grid in cash-on-hand, so consumption appears on both sides of the Euler equation: directly on the left, and through end-of-period assets $a = m - c$ inside the expectation on the right. Consumption must therefore be found by numerical root-finding at each grid point. \acrshort{egm} fixes end-of-period assets instead, which makes the expectation side known and reduces the update to the closed-form inversion in Proposition~\ref{prop-euler}. \citet{coeurdacier2020financial} extend \acrshort{ti} to Epstein-Zin preferences, treating consumption, value, and expected value as simultaneous controls; the accurate mode below follows this approach by interpolating decision rules at each candidate consumption during root-finding.

Given current guesses $(c^{(n)}, V^{(n)})$, \acrshort{ti} finds consumption at each $(m, z)$ by solving the Euler equation via bisection, then updates the value function. The certainty equivalent is again the $V$-space $\nu(a, z) = (\mathbb{E}_{z'|z}[V(m', z')^{1 -\gamma}])^{1/(1 -\gamma)}$, exactly as in \acrshort{vfi}.

\begin{algorithm}
\caption{Time Iteration for Epstein-Zin}\label{alg:ti-ez}
\begin{algorithmic}[1]
\Require Exogenous grid $\{m_i\}_{i=1}^I$, asset grid $\{a_j\}_{j=1}^J$, initial guess $(c^{(0)}, V^{(0)})$, mode $\in \{\texttt{accurate}, \texttt{fast}\}$
\Ensure Converged policy $c(\cdot, \cdot)$ and value $V(\cdot, \cdot)$
\State Given $(c^{(n)}, V^{(n)})$ from iteration $n$
\State Precompute $c^{(n)}$ and $V^{(n)}$ at $m' = Ra_j + y(z'_\ell)$ for all $(a_j, z'_\ell)$
\If{mode = \texttt{fast}}
    \State Precompute $\nu_j = (\sum_\ell \pi_{k\ell} V^{(n)}(m'_{j\ell}, z'_\ell)^{1-\gamma})^{1/(1-\gamma)}$ for all $a_j$
\EndIf
\For{each income state $z_k$ and each $m_i$}
    \State Define Euler residual $r(c)$ using $\nu$ and marginal utilities
    \If{mode = \texttt{accurate}}
        \State Compute $\nu$ by interpolating $V^{(n)}$ at $m' = R(m_i - c) + y(z')$, then taking expectation
    \Else
        \State Interpolate $\nu$ from precomputed $\{\nu_j\}$ at $a = m_i - c$
    \EndIf
    \State Solve $r(c) = 0$ via bisection to get $c^{(n+1)}(m_i, z_k)$
    \State Update $V^{(n+1)}(m_i, z_k)$ using the Bellman equation
\EndFor
\State \textbf{Iterate} until $\|c^{(n)} - c^{(n-1)}\| < \varepsilon$
\end{algorithmic}
\end{algorithm}
\section{Benchmarks}\label{sec-benchmarks}

This section reports speed and accuracy benchmarks for an infinite-horizon consumption-savings problem with stochastic income. Income follows an AR(1) process with persistence 0.95, consistent with estimates from labor market data (e.g., \citet{storesletten2004cyclical}), discretized using the \citet{tauchen1986finite} method with 10 grid points. The asset grid uses 100 exponentially spaced points with an upper bound of 20 times mean income.

\textbf{Parameters.} The calibration sets $\beta = 0.96$, $R = 1.02$, $\gamma = 10$, and \acrshort{eis} $= 1.5$ (so $\rho = 2/3$), following \citet{bansalyaron2004}. The meta-analysis of \citet{havranek2015measuring} finds that micro estimates of the \acrshort{eis} often fall below unity, though macro and asset pricing calibrations typically use higher values.\footnote{High $\gamma$ with low $\rho$ implies aversion to contemporaneous consumption risk but tolerance for intertemporal variation. Whether this reflects preferences or is a modeling device for asset pricing remains debated; the method here is agnostic on calibration.} Since $\gamma > \rho$, the agent prefers early resolution of uncertainty. The auxiliary parameter $\theta = (1 -\gamma)/(1 -\rho) = -27$.

\subsection{Measuring accuracy}

Accuracy is assessed using the normalized Euler equation error, a standard metric whose magnitude bounds the policy function error. \citet{santos2000accuracy} establishes the theoretical foundation:
\begin{equation}
\varepsilon(m, z) = \log_{10} \left| 1 - \frac{\tilde{c}(m, z)}{c(m, z)} \right|,
\end{equation}
where $\tilde{c}(m, z)$ is consumption implied by the Euler equation given the computed policy. The tables report mean (L1) and maximum (L$\infty$) errors, following \citet{maliarmaliar2014}. More negative values indicate higher accuracy; an error of -5 means the policy deviates from the Euler equation by approximately 10\textsuperscript{-5}, or 0.001\%.

Errors should be evaluated where agents spend time rather than only at arbitrary grid points, so two evaluation sets are used throughout. The grid measure $\bar{\varepsilon}_G$ averages $\varepsilon$ over interior grid points, the 10th to 90th percentiles of the grid. The ergodic measure $\bar{\varepsilon}_\pi$ averages over a sample from the stationary distribution of wealth, simulated with 10,000 agents over 500 periods and trimmed to the 5th to 95th percentiles of realized wealth.\footnote{Points near the borrowing constraint are excluded under both measures, as the Euler equation holds as an inequality there; this is standard practice following \citet{santos2000accuracy}.} Evaluating on a simulated sample from the ergodic set follows \citet{juddmaliarmaliar2011numerically}; \citet{maliarmaliar2014} survey accuracy measures of this type. It weights accuracy by economic relevance: errors at rarely visited wealth levels matter less. For this calibration the stationary distribution concentrates in the lower wealth range (median $m \approx 3$ versus grid maximum of 20), reflecting moderate impatience.

Table~\ref{tab:euler-methods} compares the three methods under both measures in fast mode. \acrshort{egm} yields similar error statistics under $\bar{\varepsilon}_G$ and $\bar{\varepsilon}_\pi$, so its solution is accurate where agents spend time. \acrshort{ti} improves under $\bar{\varepsilon}_\pi$ (-3.6 against -3.2), which suggests its errors concentrate at high wealth levels that agents rarely visit, while \acrshort{vfi} posts the worst max error of the three methods under $\bar{\varepsilon}_\pi$ (-2.3): optimization-based search is least precise at low wealth levels, where the policy function has the most curvature and agents spend the most time. \acrshort{egm} achieves the best accuracy under both measures because it satisfies the Euler equation by construction at the endogenous grid points; interpolation introduces error, but linear interpolation on a fine grid keeps it small, whereas \acrshort{ti} and \acrshort{vfi} compound interpolation error with numerical search. Unless a table notes otherwise, reported errors are the ergodic measure $\bar{\varepsilon}_\pi$.

\begin{table}[ht]
\centering
\caption{Euler errors by evaluation method ($\log_{10}$)}
\label{tab:euler-methods}
\begin{tabular}{@{}lrrrr@{}}
\toprule
 & \multicolumn{2}{c}{$\bar{\varepsilon}_G$ (grid)} & \multicolumn{2}{c}{$\bar{\varepsilon}_\pi$ (ergodic)} \\
\cmidrule(lr){2-3} \cmidrule(lr){4-5}
Method & Mean & Max & Mean & Max \\
\midrule
EZ-EGM & $-4.8$ & $-3.5$ & $-4.8$ & $-3.3$ \\
TI & $-3.2$ & $-2.7$ & $-3.6$ & $-2.7$ \\
VFI & $-3.2$ & $-2.2$ & $-3.3$ & $-2.3$ \\
\bottomrule
\end{tabular}

{\footnotesize \textit{Note:} The grid measure $\bar{\varepsilon}_G$ evaluates at 10th-90th percentiles of the grid ($m \in [0.4, 16.5]$); the ergodic measure $\bar{\varepsilon}_\pi$ at 5th-95th percentiles of simulated wealth ($m \in [0.7, 9.6]$), with median wealth $\approx 3$. Errors exclude constrained points where the Euler equation holds as inequality. All methods use fast mode (certainty equivalents precomputed on the asset grid).}
\end{table}
\subsection{Speed}

Table~\ref{tab:speed} reports times and Euler errors for the three methods at fast settings. \citet{rendahl2015inequality} proves that \acrshort{ti} and \acrshort{vfi} converge to the same solution under standard conditions, so the comparison is one of computational efficiency.

\begin{table}[ht]
\centering
\caption{Speed comparison (fast modes)}
\label{tab:speed}
\begin{tabular}{@{}lrrr@{}}
\toprule
Method & Time (ms) & Policy Iters & Euler Error \\
\midrule
EZ-EGM & 13 & 143 & $-4.8$ \\
TI (fast) & 219 & 140 & $-3.6$ \\
VFI (fast) & 758 & 203 & $-3.3$ \\
\bottomrule
\end{tabular}

{\footnotesize \textit{Note:} Euler error is the ergodic mean $\bar{\varepsilon}_\pi$ in $\log_{10}$ units (more negative = more accurate). Fast modes precompute certainty equivalents on the asset grid. Timings on CPU; results vary by hardware. All methods use JAX.}
\end{table}
\acrshort{ez}-\acrshort{egm}'s speed advantage comes from eliminating numerical search: \acrshort{ti} requires bisection at every grid point and \acrshort{vfi} requires golden-section search, while \acrshort{ez}-\acrshort{egm} bypasses both through analytic Euler inversion. At equal grid size (100 points), \acrshort{ez}-\acrshort{egm} runs roughly 60 times faster than \acrshort{vfi} and 17 times faster than \acrshort{ti}, and is more than an order of magnitude more accurate than either.

\subsection{Accuracy}

Fast modes trade accuracy for speed by interpolating a precomputed $\nu$; accurate modes compute $\nu$ exactly at every candidate during search. Table~\ref{tab:accuracy} reports the accurate-mode comparison.

\begin{table}[ht]
\centering
\caption{Accuracy comparison (accurate modes)}
\label{tab:accuracy}
\begin{tabular}{@{}lrrrr@{}}
\toprule
Method & Time (ms) & Policy Iters & Mean (L1) & Max (L$\infty$) \\
\midrule
EZ-EGM & 13 & 143 & $-4.8$ & $-3.3$ \\
TI (accurate) & 1893 & 143 & $-4.8$ & $-3.3$ \\
VFI (accurate) & 5393 & 203 & $-3.4$ & $-2.2$ \\
\bottomrule
\end{tabular}

{\footnotesize \textit{Note:} Errors are $\bar{\varepsilon}_\pi$ (ergodic). Accurate modes compute $\nu$ exactly during search. TI-accurate matches EGM accuracy but is 145$\times$ slower. VFI-accurate is 415$\times$ slower yet less accurate.}
\end{table}
Table~\ref{tab:accuracy} shows that \acrshort{ti} matches \acrshort{egm}'s accuracy (mean error -4.8) only when $\nu$ is computed exactly during search, which makes it 145 times slower. \acrshort{vfi}-accurate is 415 times slower than \acrshort{egm} and still 1.4 orders of magnitude less accurate (-3.4 versus -4.8). \acrshort{egm} and \acrshort{ti} work with the Euler equation directly while \acrshort{vfi} optimizes the Bellman equation; Euler-based methods achieve higher precision when the certainty equivalent is computed exactly. \acrshort{ti} gets there by bisection, \acrshort{egm} by analytic inversion, and the analytic path is what keeps \acrshort{egm} cheap.

\textbf{Economic significance.} Statistical accuracy in $\log_{10}$ Euler error translates into a small welfare cost. The consumption-equivalent welfare loss, the permanent consumption an agent would sacrifice to avoid using the approximate policy, is below 0.1\% under \acrshort{egm}.\footnote{Computed by simulating 20,000 agents under the approximate and high-accuracy policies with identical shocks, evaluating realized states with the high-accuracy value function in both cases. Because the Epstein-Zin aggregator is homogeneous of degree one in consumption, scaling a consumption plan by $1+\lambda$ scales value by $1+\lambda$, so the welfare cost is one minus the ratio of the two ergodic average values.} Numerical error from \acrshort{ez}-\acrshort{egm} is therefore an order of magnitude smaller than typical estimation uncertainty in this class of models, and small enough that the choice between policies is driven by the economics rather than by the solver.

\begin{figure}[!htbp]
\centering
\includegraphics[width=0.7\linewidth]{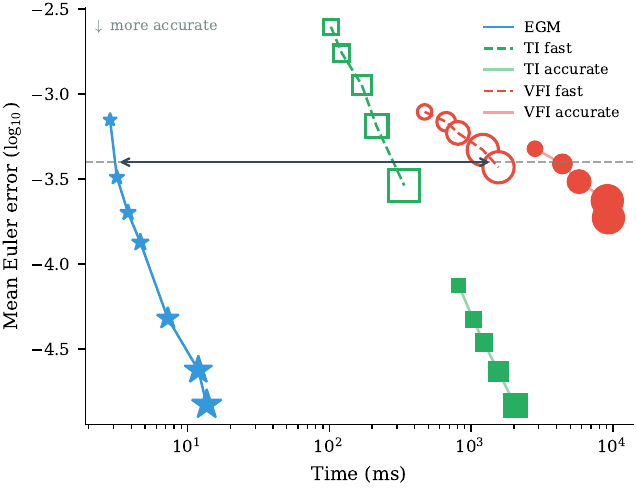}
\caption[]{Speed-accuracy Pareto frontier across grid sizes (more negative is more accurate, as in every table). Errors are the grid measure $\bar{\varepsilon}_G$. Marker size shows grid resolution $n$. The dashed line marks a reference accuracy level; \acrshort{egm} achieves it with small grids while \acrshort{vfi} and \acrshort{ti} require an order of magnitude more grid points to match it. The horizontal arrow measures the equal-accuracy gap at that level, where \acrshort{egm} is several hundred times faster than \acrshort{vfi}.}
\label{fig-pareto}
\end{figure}

\textbf{Equal-accuracy comparison.} A more informative comparison holds accuracy constant. Figure~\ref{fig-pareto} shows that to match \acrshort{egm}'s Euler errors with 15-20 grid points, \acrshort{vfi} requires 50-300 points; the resulting speedup is 166-744 times. This comparison isolates computational efficiency: both methods deliver the same solution quality, but \acrshort{egm} does so with far less computation. For applications requiring many model evaluations (estimation, uncertainty quantification, real-time decision support), such speedups translate directly into feasibility. Table~\ref{tab:equal-accuracy} reports the underlying grid sizes and times: the speedup grows with the accuracy target, and matching a mean error near -3.6 takes \acrshort{vfi} 300 grid points and 2.5 seconds against \acrshort{egm}'s 20 points and 3.4 ms.

\begin{table}[ht]
\centering
\caption{Speed comparison at equal accuracy}
\label{tab:equal-accuracy}
\begin{tabular}{@{}rrrrrr@{}}
\toprule
VFI $n$ & EGM $n$ & Mean Error & VFI (ms) & EGM (ms) & Speedup \\
\midrule
50 & 15 & $-3.1$ & 497 & 3.0 & 166$\times$ \\
100 & 15 & $-3.2$ & 869 & 3.0 & 291$\times$ \\
150 & 20 & $-3.3$ & 1280 & 3.4 & 381$\times$ \\
200 & 20 & $-3.4$ & 1684 & 3.4 & 502$\times$ \\
300 & 20 & $-3.6$ & 2496 & 3.4 & 744$\times$ \\
\bottomrule
\end{tabular}

{\footnotesize \textit{Note:} Mean Error is VFI's grid-measure ($\bar{\varepsilon}_G$) mean Euler error at the listed grid size; each row pairs it with the benchmarked EGM grid size whose mean error is closest. EGM's own mean errors are $-3.15$ at $n = 15$ and $-3.49$ at $n = 20$. Speedups are computed from unrounded times, so they differ slightly from ratios of the rounded columns.}
\end{table}
These results use $\beta R = 0.979 < 1$, so wealth converges to a finite target. More impatient agents ($\beta R \approx 0.92$) have lower target wealth but similar accuracy. For very patient agents ($\beta R \geq 1$) target wealth diverges, no fixed grid contains the wealth distribution, and the accuracy comparison in that regime warrants separate investigation; the same containment logic governs the grid choice in the~robustness~analysis.

\begin{figure}[!htbp]
\centering
\includegraphics[width=0.7\linewidth]{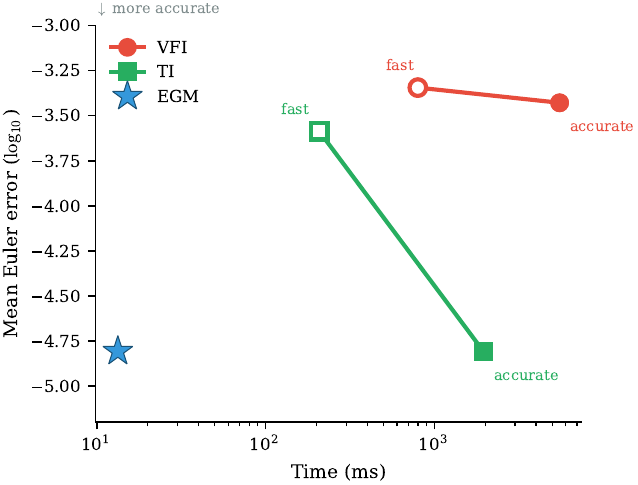}
\caption[]{Speed-accuracy tradeoff (more negative is more accurate, as in Figure~\ref{fig-pareto} and every table). \acrshort{vfi} and \acrshort{ti} trade speed for accuracy between their fast modes (open markers) and accurate modes (filled markers). \acrshort{egm} sidesteps the tradeoff: a single point in the fast, accurate corner at the bottom left achieves both.}
\label{fig-tradeoff}
\end{figure}

\textbf{Speed-accuracy tradeoff.} Time iteration is a middle ground between \acrshort{egm} and \acrshort{vfi}: like \acrshort{egm} it works with the Euler equation; like \acrshort{vfi} it requires numerical optimization. Figure~\ref{fig-tradeoff} illustrates the tradeoff: computing $\nu$ exactly at each candidate is accurate but slow (Table~\ref{tab:accuracy}); precomputing $\nu$ on the asset grid and interpolating is fast but introduces approximation error (Table~\ref{tab:speed}). The fast modes in Table~\ref{tab:speed} are the practical implementations. \acrshort{egm} has no analogous choice to make: its endogenous grid is built from the Euler equation, so it achieves accurate-mode precision at fast-mode speed.

\begin{figure}[!htbp]
\centering
\includegraphics[width=0.7\linewidth]{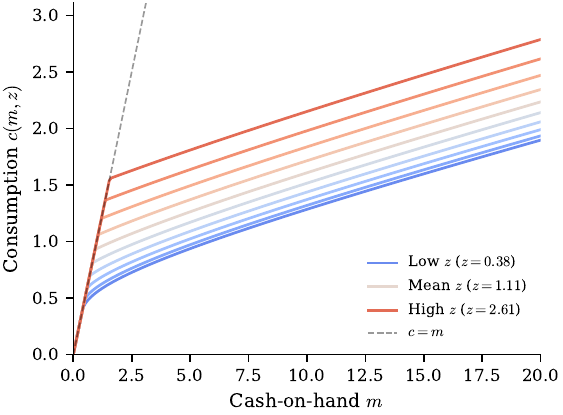}
\caption[]{Converged consumption policy $c(m, z)$ across income states. Consumption rises with income at each wealth level (red marks high $z$), and the only kink is at the borrowing constraint where $c = m$.}
\label{fig-policy}
\end{figure}

The policy these speedups deliver is itself well-behaved. Figure~\ref{fig-policy} plots the converged consumption function: it is smooth across income states and kinks only at the borrowing constraint, with no interpolation artifacts.

\textbf{Howard acceleration.} The baseline algorithm performs one policy update per value update. To accelerate convergence, fix the policy $c(\cdot, \cdot)$ and iterate on $W$ alone:
\begin{equation}
W^{(k+1)}(m, z) = (1-\beta) c(m,z)^{1-\rho} + \beta \mu^{(k)}(m-c, z),
\end{equation}
where $\mu^{(k)}$ uses $W^{(k)}$. After $K$ value iterations, the policy is updated via \acrshort{egm}. For \acrshort{ez}-\acrshort{egm}, $K=1$ is fastest (Table~\ref{tab:howard-egm}): additional value iterations reduce policy iterations from 143 at $K=1$ to roughly 100, where they plateau, but increase total time because the \acrshort{egm} policy step is already cheap and accurate.

\begin{table}[ht]
\centering
\caption{EZ-EGM: Effect of Howard acceleration parameter $K$}
\label{tab:howard-egm}
\begin{tabular}{@{}rrrcc@{}}
\toprule
$K$ & Time (ms) & Policy Iters & Euler Mean & Euler Max \\
\midrule
1 & 14 & 143 & $-4.8$ & $-3.5$ \\
2 & 40 & 101 & $-4.9$ & $-3.5$ \\
3 & 60 & 100 & $-5.0$ & $-3.5$ \\
4 & 78 & 100 & $-5.0$ & $-3.5$ \\
5 & 98 & 101 & $-5.0$ & $-3.5$ \\
\bottomrule
\end{tabular}

{\footnotesize \textit{Note:} Errors are $\bar{\varepsilon}_G$; the $\bar{\varepsilon}_\pi$ evaluation used in Tables~\ref{tab:speed} and~\ref{tab:accuracy} yields mean $-4.8$ and max $-3.3$ at $K=1$.}
\end{table}
The slight accuracy improvement from $K=1$ to $K=2$ (-4.8 to -4.9) does not justify the nearly threefold time increase for most applications, but $K=2$ remains faster than all \acrshort{vfi} and \acrshort{ti} configurations. Across the comparison methods, the optimal $K$ differs sharply, and the contrast is instructive.\footnote{Iteration accounting: for \acrshort{ti}, each policy iteration is one bisection solve at every grid point plus up to $K$ value updates; for \acrshort{vfi}, one numerical optimization over the consumption grid plus up to $K$ value updates. Both methods terminate the inner loop early if the value function converges before $K$ updates.} The Howard sweeps come from a separate timing run, so their $K = 1$ rows differ from Table~\ref{tab:speed} and Table~\ref{tab:accuracy} by run-to-run timing variation of up to ten percent, largest in relative terms for the fastest configurations; iteration counts and Euler errors are deterministic and agree exactly.

\acrshort{vfi} benefits substantially from Howard acceleration in both modes, with optimal $K \approx 30$ cutting time by 7x in fast mode and 16x in accurate mode (Table~\ref{tab:howard-vfi}). \acrshort{vfi}-accurate benefits more because each policy optimization is more expensive when $\nu$ is computed exactly: golden-section search is the bottleneck, so reducing policy iterations helps most when each policy step is dear. At $K = 30$, \acrshort{vfi}-accurate (354 ms) is faster than \acrshort{vfi}-fast at $K = 1$ (763 ms), though still less accurate.

\begin{table}[ht]
\centering
\caption{VFI: Effect of Howard acceleration parameter $K$}
\label{tab:howard-vfi}
\begin{tabular}{@{}rrrrr@{}}
\toprule
 & \multicolumn{2}{c}{Fast} & \multicolumn{2}{c}{Accurate} \\
\cmidrule(lr){2-3} \cmidrule(lr){4-5}
$K$ & Time (ms) & Policy Iters & Time (ms) & Policy Iters \\
\midrule
1 & 763 & 203 & 5527 & 203 \\
10 & 155 & 28 & 798 & 28 \\
20 & 118 & 16 & 483 & 16 \\
30 & 111 & 12 & 354 & 11 \\
40 & 122 & 11 & 376 & 11 \\
50 & 141 & 11 & 430 & 12 \\
\bottomrule
\end{tabular}

{\footnotesize \textit{Note:} Euler errors ($\bar{\varepsilon}_G$ mean/max) are essentially invariant to $K$: $-3.2$/$-2.2$ in fast mode (max $-2.1$ for $K \geq 10$), $-3.5$/$-2.3$ in accurate mode.}
\end{table}
\acrshort{ti} gains less from Howard acceleration (Table~\ref{tab:howard-ti}). \acrshort{ti}-fast improves only modestly, from 201 ms at $K = 1$ to about 177 ms for $K$ between 3 and 5, because bisection already produces accurate policy updates, so the policy step rather than the value step dominates the cost. \acrshort{ti}-accurate improves monotonically through $K = 5$ (1904 ms down to 987 ms): each avoided policy iteration saves a full pass of exact-$\nu$ bisection, which is where accurate mode spends its time.

\begin{table}[ht]
\centering
\caption{TI: Effect of Howard acceleration parameter $K$}
\label{tab:howard-ti}
\begin{tabular}{@{}rrrrr@{}}
\toprule
 & \multicolumn{2}{c}{Fast} & \multicolumn{2}{c}{Accurate} \\
\cmidrule(lr){2-3} \cmidrule(lr){4-5}
$K$ & Time (ms) & Policy Iters & Time (ms) & Policy Iters \\
\midrule
1 & 201 & 140 & 1904 & 143 \\
2 & 183 & 100 & 1352 & 100 \\
3 & 177 & 88 & 1197 & 87 \\
4 & 177 & 80 & 1101 & 79 \\
5 & 176 & 71 & 987 & 70 \\
\bottomrule
\end{tabular}

{\footnotesize \textit{Note:} Euler errors ($\bar{\varepsilon}_G$ mean/max) in fast mode are $-3.2$/$-2.7$ at every $K$; in accurate mode, $-4.8$/$-3.5$ at $K=1$ and $-4.9$/$-3.5$ for $K \geq 2$.}
\end{table}
The contrast traces back to the quality of each method's policy step. Golden-section search produces noisy policy updates, so \acrshort{vfi} needs many value iterations to stabilize each one and benefits from large $K$. Bisection produces accurate policies, so \acrshort{ti} gains comparatively little from additional value iterations. \acrshort{egm}'s analytic policy step is both cheap and exact, which is why its optimum is $K = 1$.

\subsection{Robustness to $\rho$}\label{sec-robustness}

The baseline uses $\rho = 2/3$ (\acrshort{eis} $= 1.5$), but the algorithm applies equally to $\rho > 1$ (\acrshort{eis} $< 1$), the region consistent with the micro \acrshort{eis} estimates noted above. The calibration satisfies the impatience conditions of buffer-stock theory \citep{carroll1997buffer} at every $\rho$ considered: with stationary income the absolute and growth impatience conditions coincide and require $(\beta R)^{1/\rho} < 1$, return impatience requires $(\beta R)^{1/\rho} < R$, and both hold at $\beta R = 0.979$. A stationary wealth distribution therefore exists throughout. Its support widens as the \acrshort{eis} falls (median simulated wealth rises from about 3 at the baseline to about 13 at $\rho = 3$), so the grid for this exercise extends to an upper bound of 100 rather than the baseline 20, large enough to contain the stationary distribution at every $\rho$. Table~\ref{tab:rho-robustness} reports results for $\rho \in \{0.5, 0.9, 1.1, 1.5, 2, 3\}$ with $\gamma = 10$ fixed: ergodic-weighted mean Euler errors stay between -4.6 and -4.7, max errors between -3.66 and -4.09, and iteration counts rise modestly with $\rho$, from 158 at $\rho = 0.5$ to 230 at $\rho = 3$, against 143 at the baseline.

\begin{figure}[!htbp]
\centering
\includegraphics[width=0.7\linewidth]{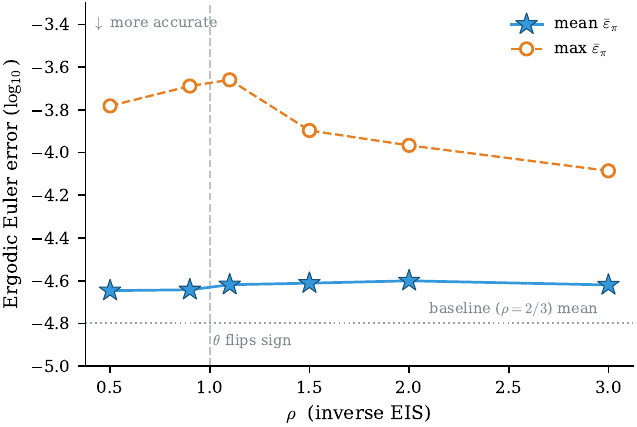}
\caption[]{Ergodic Euler error across the \acrshort{eis}, with $\gamma = 10$ fixed. Mean (stars) and max (circles) errors stay within a narrow band as $\rho$ crosses 1, where the auxiliary parameter $\theta = (1 -\gamma)/(1 -\rho)$ changes sign (dashed line). The algorithm is unchanged on either side.}
\label{fig-robustness}
\end{figure}

\begin{table}[ht]
\centering
\caption{EZ-EGM across the EIS}
\label{tab:rho-robustness}
\begin{tabular}{@{}rrrrrr@{}}
\toprule
$\rho$ & EIS & $\theta$ & Iters & Mean & Max \\
\midrule
0.5 & 2.00 & $-18.0$ & 158 & $-4.65$ & $-3.78$ \\
0.9 & 1.11 & $-90.0$ & 165 & $-4.64$ & $-3.69$ \\
1.1 & 0.91 & $90.0$ & 178 & $-4.62$ & $-3.66$ \\
1.5 & 0.67 & $18.0$ & 201 & $-4.61$ & $-3.90$ \\
2.0 & 0.50 & $9.0$ & 223 & $-4.60$ & $-3.97$ \\
3.0 & 0.33 & $4.5$ & 230 & $-4.62$ & $-4.09$ \\
\bottomrule
\end{tabular}

{\footnotesize \textit{Note:} Risk aversion is held at $\gamma = 10$; other parameters as in Table~\ref{tab:speed}. The grid keeps 100 points; its upper bound is raised to 100 units of mean income (from 20 in Table~\ref{tab:speed}) so that it contains the stationary wealth distribution at every $\rho$. Mean and max are $\bar{\varepsilon}_\pi$ (ergodic) Euler errors in $\log_{10}$ units, as in Table~\ref{tab:speed}. The auxiliary parameter $\theta = (1-\gamma)/(1-\rho)$ changes sign at $\rho = 1$, but the algorithm requires no modification.}
\end{table}
Figure~\ref{fig-robustness} plots these errors against $\rho$: both the mean and the max stay essentially flat as $\rho$ sweeps through the $\theta$ sign change at $\rho = 1$, so accuracy is insensitive to the \acrshort{eis} and the method needs no retuning.

When $\rho > 1$, the transformation $W = V^{1 -\rho}$ is decreasing, so maximizing $V$ is equivalent to minimizing $W$, as (\ref{eq-bellman-w}) records. No algorithmic modification is required: the same first-order condition $\partial W / \partial c = 0$ characterizes the interior optimum in both branches, its sufficiency follows from the strict monotonicity of the transformation together with the concavity of the original problem in $c$,\footnote{Concavity propagates by induction. A power mean with exponent at most one is concave and increasing in its arguments, so if $V(\cdot, z')$ is concave for every $z'$, the certainty equivalent $\nu(a, z)$ of the~direct~approach is concave in $a$. The CES aggregator in Definition~\ref{def-model} is itself a power mean of $c$ and $\nu$ with exponent $1 -\rho \leq 1$, hence concave in $(c, a)$ jointly, and maximization over the convex budget set preserves concavity, so $V(\cdot, z)$ is concave; the terminal condition (or the concave initial guess of the~algorithm~section) starts the induction. In transformed terms the objective in (\ref{eq-bellman-w}) is therefore concave in $c$ when $\rho < 1$ and convex when $\rho > 1$, so the first-order condition identifies the interior optimum on both branches.} and \acrshort{egm} inverts the Euler equation analytically either way.

\acrshort{vfi} is unaffected because it works with $V$ directly; the CES aggregator in Definition~\ref{def-model} remains increasing in $c$ for all $\rho \neq 1$. Time iteration, like \acrshort{egm}, works with the Euler equation.

\section{Conclusion}\label{sec-conclusion}

The endogenous grid method extends to Epstein-Zin preferences through the transformation $W = V^{1 -\rho}$. Time-iteration methods can solve the Epstein-Zin Euler equation by root-finding, so the transformation is not strictly necessary, but it is structurally natural: after the transformation $\mu$ is a power mean, the quasi-arithmetic family with closed-form inverse, and recovering consumption from end-of-period assets is exactly this inversion. That inversion is the structural reason \acrshort{egm} works for Epstein-Zin in one asset without numerical search, and it produces the order-of-magnitude speed and accuracy gains documented in the~benchmark~section.

Beyond Epstein-Zin, the same observation extends to other recursive preferences whose certainty equivalent admits an analytic inverse. The limit $\rho \to 1$ yields the risk-sensitive preferences of \citet{hansensargent1995}, whose exponential certainty equivalent $\frac{1}{1 -\gamma} \log \mathbb{E}[\exp((1 -\gamma) W)]$ has generator $\varphi(w) = \exp((1 -\gamma) w)$ and admits direct log-inversion. The limit $\gamma \to 1$ yields a geometric mean (generator $\varphi(w) = \log w$), also invertible directly. In each case the \acrshort{egm} step exploits this inverse, and the rest of the algorithm follows the standard one-asset template.

Two extensions are immediate. First, the closed-form-inversion argument is local to the certainty-equivalent structure and does not depend on the dimensionality of the state space, so the same observation combines naturally with the multi-dimensional \acrshort{egm} techniques of \citet{druedahljorgensen2017} and the discrete-continuous upper envelopes of \citet{iskhakov2017endogenous}. Two-asset portfolio choice with Epstein-Zin preferences, the setting of \citet{kaplanviolante2014}, fits squarely in this combined framework. Second, life-cycle models with deterministic age effects on income or mortality risk introduce time-varying $\beta$ and $R$ in the recursion but leave the inversion in Proposition~\ref{prop-euler} unchanged; the algorithm requires only the backward induction familiar from finite-horizon \acrshort{egm}. Estimation pipelines that solve the model thousands of times during simulated method of moments or indirect inference stand to benefit most directly from the speed gains reported here.
\printglossaries

\backmatter\section*{Declarations}
\textbf{Funding.} This work was supported by the Alfred P. Sloan Foundation (Grant G-2025-79177).

\textbf{Competing interests.} The author has no relevant financial or non-financial interests to disclose.

\textbf{Author contributions.} The author is solely responsible for the conception and design of the study, the development and implementation of the algorithm, the numerical analysis, and the writing of the manuscript.

\textbf{Data and code availability.} No external datasets were analyzed or generated. All numerical results are reproducible from the publicly available code at \href{https://github.com/alanlujan91/ezegm}{https://github.com/alanlujan91/ezegm}, archived on Zenodo under DOI 10.5281/zenodo.18176983.

\textbf{Use of generative AI.} During the preparation of this work the author used Claude (Anthropic) to assist with code development and manuscript editing. After using this tool, the author reviewed and edited the content as needed and takes full responsibility for the content of the published article.
\bibliography{main}
\end{document}